\documentclass[final,5p,times,twocolumn]{elsarticle}

\usepackage{graphicx}
\usepackage{amssymb}
\usepackage{amsmath}
\journal{Nuclear Instruments and Methods in Physics Research A}

\begin{document}

\begin{frontmatter}



\title{Neutron imaging detector based on the $\mu$PIC micro-pixel chamber}


\author[ku]{J.D.~Parker\corref{cor1}}
\ead{jparker@cr.scphys.kyoto-u.ac.jp}

\author[ku]{K.~Hattori}
\author[ku]{H.~Fujioka}
\author[jaea]{M.~Harada}
\author[ku]{S.~Iwaki}
\author[ku]{S.~Kabuki}
\author[ku]{Y.~Kishimoto}
\author[ku]{H.~Kubo}
\author[ku]{S.~Kurosawa}
\author[ku]{K.~Miuchi}
\author[ku]{T.~Nagae}
\author[ku]{H.~Nishimura}
\author[jaea]{T.~Oku}
\author[ku]{T.~Sawano}
\author[jaea]{T.~Shinohara}
\author[jaea]{J.~Suzuki}
\author[ku]{A.~Takada}
\author[ku]{T.~Tanimori}
\author[ku]{K.~Ueno}

\cortext[cor1]{Corresponding author}

\address[ku]{Department of Physics, Graduate School of Science, Kyoto University, Kitashirakawa-oiwakecho, Sakyo-ku, Kyoto 606-8502, Japan}
\address[jaea]{Materials and Life Science Facility Division, Japan Atomic Energy Agency (JAEA), Tokai, Ibaraki 319-1195, Japan}

\begin{abstract}

We have developed a prototype time-resolved neutron imaging detector employing
the micro-pixel chamber ($\mu$PIC), a micro-pattern gaseous detector,
coupled with a field programmable gate array-based data acquisition system
for applications in neutron radiography at high-intensity neutron sources.
The prototype system,
with an active area of $10 \times 10$~cm$^2$ and operated at a gas
pressure of 2~atm,
measures both the energy deposition (via time-over-threshold) 
and 3-dimensional track of each neutron-induced event,
allowing the reconstruction of the neutron
interaction point with improved accuracy.
Using a simple position reconstruction algorithm, 
a spatial resolution of $349 \pm 36$~$\mu$m was achieved,
with further improvement expected.
The detailed tracking allows strong rejection of background 
gamma-rays,
resulting in an effective gamma sensitivity of 10$^{-12}$ or less,
coupled with stable, robust neutron identification.
The detector also features a time resolution of 0.6~$\mu$s.

\end{abstract}


\end{frontmatter}

% \linenumbers

\section{Introduction}
\label{sec:intro}

The micro-pixel chamber ($\mu$PIC)~\cite{ochi01},
a type of micro-pattern gaseous detector~\cite{oed88},
has been developed in our group at Kyoto University.
The $\mu$PIC features a 400-$\mu$m pitch with 2-dimensional strip readout
coupled with a fast, compact FPGA (Field Programmable Gate Array)-based
data acquisition system.
The 2-dimensional measurement of the $\mu$PIC can be extended to three
dimensions using
an arrangement known generally as a time projection chamber (TPC).
Detectors incorporating the $\mu$PIC have already been successfully 
developed for applications ranging from MeV gamma-ray astronomy~\cite{takada05} 
to medical imaging~\cite{kabuki07}
to small-angle X-ray scattering~\cite{hattori09}.
By adding $^3$He to the gas mixture, such a system becomes an effective
thermal neutron area-detector, well suited to neutron radiography
at high-intensity neutron sources.

State of the art neutron radiography measurements
%important tools for the study of materials on the scale of mm to nm, 
require detectors capable of high
spatial resolutions from several hundreds to tens of micrometers
and a wide dynamic range for good image contrast.
Commonly used neutron imaging detectors satisfying these requirements include
imaging plates~\cite{nip_kobayashi99} 
and CCD (charge-coupled device)-based systems~\cite{ccd_pleinert97}.
Imaging plates (IP) are capable of spatial resolutions down to 25~$\mu$m,
while CCD systems generally achieve resolutions around 100~$\mu$m. 
Each of these detector types also features a wide dynamic range
(on the order of 10$^5$)
and good linearity in response to intensity.
On the other hand,
read-out times can be long,
particularly in the case of IPs,
which must be removed from the beam and scanned after each measurement.
Another drawback of these integrating-type detectors is a background
proportional to measurement time, with no means of separating the neutron-
and gamma-induced events.

To take full advantage of the new generation of high-intensity, pulsed spallation 
neutron sources
and time-resolved radiographic techniques such as Bragg-edge transmission and
resonance imaging,
a detector which also possesses
good time resolution 
and good discrimination against gammas is essential.
At pulsed sources, the energy of an incident neutron is determined
by measuring the time between the initiation of the pulse and
the neutron's arrival at the detector
(referred to as the time-of-flight, or TOF).
This requires a time resolution
generally on the order of microseconds.
IPs are not suitable for measuring neutron TOF since 
typical exposure times are on the order of seconds or greater.
In the case of CCDs, 
it is possible to achieve
frame rates on the order of kHz (down to $\sim$0.1~ms per frame)~\cite{fastccd},
but the ability to measure neutron TOF remains limited.
The gamma sensitivity of the IPs and CCDs is also an issue, since gammas are 
produced in copious amounts at spallation sources.

Our $\mu$PIC-based detector was able to overcome the limitations
of IP and CCD systems,
while maintaining many of their good characteristics.
This stems from the fact that
the $\mu$PIC is a counting-type detector (i.e., each neutron interaction
is measured individually, or counted).
Our prototype detector was able to assign a time for each event with
a resolution down to 0.6~$\mu$s,
which is adequate for separating neutrons by TOF at pulsed sources.
Also, examination of the properties of each event (track length, energy
deposition, etc.) provided a powerful means for rejecting background
on an event-by-event basis,
resulting in an effective gamma sensitivity of ${\scriptstyle \lesssim}$10$^{-12}$
while maintaining robust, stable neutron identification under
varying detector conditions.
The detailed event-by-event tracking also allowed our prototype
neutron detector to achieve a spatial resolution of $349 \pm 36$~$\mu$m.
The detector has a neutron detection efficiency of 
up to 35\% for thermal neutrons with a peak efficiency of 75\%
for cold neutrons (from simulation).

For a similar $\mu$PIC-based X-ray imaging detector,
a dynamic range comparable to IPs and CCD systems of 10$^5$ was 
observed~\cite{hattori09},
and with our prototype detector,
good linearity versus incident intensity was confirmed for
data rates up to 9~MHz 
(corresponding to $\sim$$1.5 \times 10^5$~neutrons/s)~\cite{parker10}.
[The data rate in Ref.~\cite{parker10} (4.5~MHz) is 
quoted {\em per memory port} and should be multiplied by two for the
total data rate.]
The maximum data rate was limited by the data acquisition hardware
and can be increased by as much as a factor of 10 or more 
(as described in Sec.~\ref{sec:daq}).
Furthermore, the $\mu$PIC, manufactured via inexpensive printed-circuit
board technology, allows freedom to tailor the size and shape
for a particular application,
up to a maximum size of $30 \times 30$~cm$^2$.
Larger areas can be accommodated by tiling detectors,
with dead space between the individual $\mu$PICs as small as a few millimeters
and, since the underlying anode structure remains unchanged, 
suffering no degradation in the operating characteristics with increasing area.
Thus, our $\mu$PIC-based detector can combine a moderate spatial 
resolution, good time resolution, low gamma sensitivity, and
large detection area with a wide dynamic range and high rate capability.

In the present paper, we describe the prototype detector and 
FPGA-based data acquisition system, as well as the analysis method for the
resulting data.
This is followed by a discussion of the gamma rejection capability,
spatial resolution, and long-term operation of the detector.

\section{Neutron imaging detector based on the $\mu$PIC}
\label{sec:proto}

\subsection{Detection method}
\label{sec:method}

Neutron detection was achieved via the absorption reaction, 
$^3$He($n,p$)$^3$H, which has a large cross-section for 
thermal neutrons (5,330~barns at $E_n \simeq 25.3$~meV)
and a Q-value of 764~keV.
Owing to this large energy deposition and overall track lengths 
around 1~cm in the gas of the detector,
the resulting proton-triton pairs were easily detected by the $\mu$PIC
with gas gains as low as a few hundred.
(For comparison, note that the $\mu$PIC has achieved stable operation for
gas gains up to 6000~\cite{bouianov05}.)
As typical neutron energies used in radiography measurements 
are negligible compared to the energy released in the detection reaction,
the neutron interactions result in
events with similar track lengths and energy deposition,
allowing for easy separation from most backgrounds.

The negligible energy of the incident neutron also
means the proton and triton emerge effectively back-to-back,
but as the range of the proton is $\sim$3 times greater than that of 
the heavier triton (from simple kinematics),
separation of the two particles is essential for an accurate determination
of the neutron interaction point.
By measuring the energy deposition {\em along} the track,
the proton and triton can often be cleanly separated,
facilitating a substantial improvement in spatial resolution.
The feasibility of detecting neutrons with the $\mu$PIC and distinguishing
protons and tritons via energy deposition was demonstrated 
in a previous experiment~\cite{tanimori04}.
In the prototype system, the energy deposit at each hit point was
estimated using the time-over-threshold method,
which measures the width (in time) of an analog pulse at a given threshold 
voltage~\cite{hattori12}.
Measuring the energy deposition in this way also gave another handle for 
rejecting background events.

\subsection{Design of the prototype detector}
\label{sec:design}

\begin{figure}[ht]
\centering
\includegraphics[width=8.5 cm,clip]{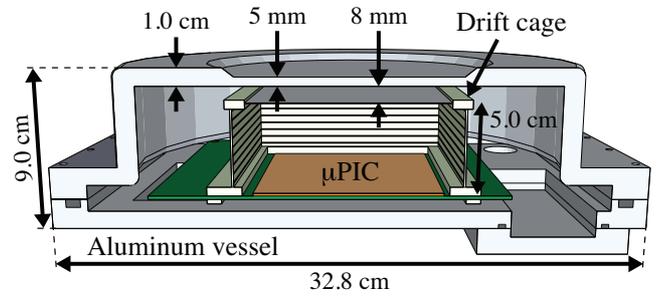}
\caption{\label{fig:nid} Diagram showing a cut-away of the 
neutron imaging detector prototype with a 5-cm tall drift cage.
}
\end{figure}

\begin{figure}[ht]
\centering
\includegraphics[width=7.5 cm,clip]{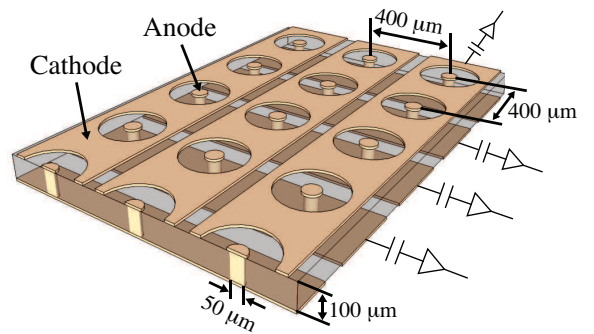}
\caption{\label{fig:muschem} Schematic diagram of the $\mu$PIC.
Anode pixels are connected on the backside by copper strips and,
combined with the orthogonal cathode strips on the top-side,
produce a 2-dimensional readout. }
\end{figure}

For the current study, 
we constructed a dedicated neutron imaging detector 
prototype (shown in Fig.~\ref{fig:nid}) 
based on a $\mu$PIC with an active area of $10 \times 10$~cm$^2$.
The $\mu$PIC,
manufactured by Dai Nippon Printing Co., Ltd., Japan,
consists of a 100-$\mu$m thick polyimide substrate with 
copper cathode strips and anode pixels at a pitch of 400~$\mu$m.
The layout of the anodes and cathodes are shown in Fig.~\ref{fig:muschem}.
Strips connecting the anodes on the underside of the substrate
combine with the orthogonal cathode strips 
to produce a two-dimensional strip read-out.
(See Ref.~\cite{ochi01} for a detailed description of the $\mu$PIC.)

The active volume of the detector 
was created using a rectangular drift cage affixed above the $\mu$PIC,
with sides constructed from G10 glass-reinforced epoxy laminate 
and topped with a drift plane consisting
of 0.3-mm thick aluminum sheet. 
The sides of the drift cage 
were etched with 0.5-mm copper lines at a 5-mm pitch, and the drift plane 
and each successive copper line were connected 
via a chain of 50~M$\Omega$ resistors.
Electric field strengths
ranged from 250 $\sim$ 800~V/cm, the optimum value
depending on the particular gas mixture and pressure.

The $\mu$PIC and drift cage were contained within a sealed aluminum
vessel with a 5-mm thick entrance window,
outer dimensions of 32.8~cm $\times$ 32.8~cm $\times$ 9.0~cm, 
and a total gas volume of 2.5~L.
The vessel could accommodate drift-cage heights up to 5~cm and
was capable of withstanding gas pressures of over 2~atm.
A 5-cm drift cage was used in the measurement of the spatial resolution
described in Sec.~\ref{sec:pres},
while the remaining data presented in this paper was taken with a 2.5-cm
drift cage.
The motivation for reducing the height of the drift cage will be elucidated
in Sec.~\ref{sec:pres}.

The filling gas consisted of Ar-C$_2$H$_6$-$^3$He (63:7:30) at 2~atm.
Our experience with $\mu$PIC-based detectors has shown that
the above stopping gas and quencher (Ar-C$_2$H$_6$ at 9:1)
provide good gain and stability characteristics.
For our initial detector study, 
we simply added $^3$He to this standard
mixture in a ratio providing adequate neutron detection efficiency.
The pressure of 2~atm was chosen for increased efficiency and stopping power,
while allowing the vessel and entrance window to remain relatively thin.
The required drift field, 800~V/cm, was determined using 
the MAGBOLTZ program~\cite{magboltz}.
Optimization of the gas mixture for detection efficiency, 
spatial resolution, etc., will be the subject of a future study.

To estimate the neutron detection efficiency, and aide in our study of 
other detector properties,
a simulation of the prototype detector was created using the GEANT4 software
toolkit~\cite{geant4},
and included the relevant physical processes
(scattering, absorption, multiple scattering, electron diffusion, etc.).
The simulation also emulated the response of the $\mu$PIC and data acquisition
hardware to produce data identical in structure to the real system,
which was then analyzed with the same software as the real data
(described in Sec.~\ref{sec:ana}).
The GEANT4 simulation included the detector structures shown in 
Fig.~\ref{fig:nid} with the gas mixture and drift field described above.

\begin{figure}[ht]
\centering
\includegraphics[width=7.5 cm,clip]{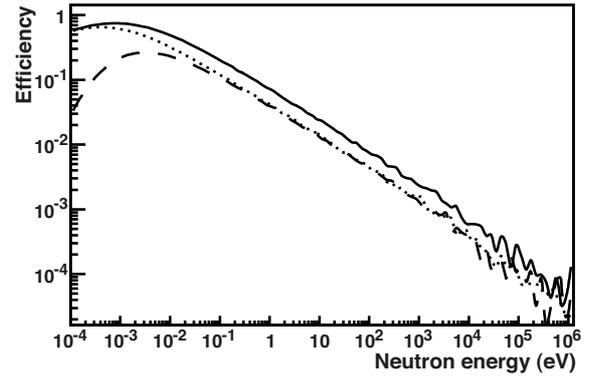}
\caption{\label{fig:neff} Expected neutron detection efficiency.
Detection efficiency versus neutron energy as determined by a GEANT4
simulation is shown for the prototype detector with a 5-cm (solid line) 
and 2.5-cm (dashed line) drift height.
The dotted line shows the efficiency for the 2.5-cm drift height when
the window-to-drift-plane distance is equal to the 5-cm case. 
}
\end{figure}

Fig.~\ref{fig:neff} shows the calculated detection efficiencies
as a function of neutron energy for the 5-cm (solid line)
and 2.5-cm (dashed line) drift cages.
The 5-cm drift cage gives a value of 35\% for thermal neutrons 
($E_n \simeq 25.3$~meV) and a peak efficiency of 75\% at 0.7~meV.
A drift height of 2.5~cm effectively halves the efficiency,
resulting in a value of 18\% for thermal neutrons. 
The large drop in efficiency below a neutron energy of about 1~meV
in the 2.5-cm case is a result of absorption in
the space between the entrance window and drift plane,
resulting in an attenuation of the incident beam
that increases with decreasing neutron energy.
For comparison, the attenuation for the 5-cm drift cage with an 
8-mm window-to-drift-plane gap is 13\% at 0.7~meV,
while that of the 2.5-cm cage (3.3-cm gap) is 74\% 
at the same neutron energy.
Reducing the height of the vessel to achieve
the same window-to-drift-plane spacing as the 5-cm case
would result in a peak efficiency of 65\% at 0.35~meV
(dotted line in Fig.~\ref{fig:neff}).

\subsection{Data acquisition system}
\label{sec:daq}

In the data acquisition system (DAQ), the charge
deposited on each anode and cathode strip, 
after being amplified by the $\mu$PIC,
was read out through ASICs (Application-Specific Integrated Circuits) containing
Amplifier-Shaper-Discriminators (ASDs)~\cite{orito04}.
The ASDs produce an LVDS (Low Voltage Differential Signaling) 
digital level whenever the analog signal from a 
strip crosses a common threshold voltage.
These digitized signals were then sent to an FPGA encoder module 
to be synchronized with a 100-MHz internal clock and 
the position and time
encoded into a 32-bit data word~\cite{kubo05}.
The encoded data was then sent over two 50-MHz parallel transfer lines
to a VME memory module with a total capacity of 64~MB (16.8~Mwords), 
where they were read out to a PC and saved for further analysis.

This system was originally developed for use with low-rate $\mu$PIC-based
gamma imaging detectors.
In high-rate environments, the encoder-to-VME-memory transfer creates a
bottleneck, limiting the maximum data rate to $\sim$10~MHz
(1~Hz $=$ 1 word/s).
This maximum data rate could be increased an order of magnitude
by employing multiple encoder modules transferring data
directly to PC over Gigabit Ethernet.
For example,
using four such encoder modules (each handling 128 strips) would
give an effective transfer rate of 4~Gb/s. 
This is equivalent to a data rate of 125~MHz, 
or a 12.5-fold increase.

\begin{figure}[ht]
\centering
\includegraphics[width=6. cm,clip]{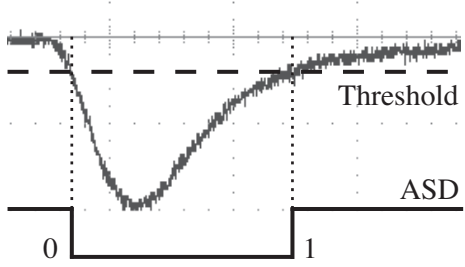}
\caption{\label{fig:pulse} Illustration of FPGA encoder logic.
An analog signal from the $\mu$PIC (oscilloscope trace) is superimposed
with a representation of the ASD threshold level (dashed line) 
and ASD output signal (solid black line).
The start (0) and end (1) times of the ASD signal are encoded in 
separate data words. 
}
\end{figure}

The logic of the FPGA encoder was
a key to the successful operation of our prototype detector.
The version of the encoder logic used here,
which simply encoded and streamed all incoming data,
was designed for use at high-intensity 
X-ray or neutron sources where an event-by-event trigger is not practical.
Additionally,
each pulse from the $\mu$PIC produced not one, but two data words,
one at the leading edge of the LVDS signal and
a second at the trailing edge,
marking each time the analog signal crossed the ASD threshold
(as illustrated in Fig.~\ref{fig:pulse}).
The resulting data words contained the strip number, the time
relative to the internal clock of the encoder, and
an {\em edge} bit used to indicate a leading or trailing edge.
The time between these hit pairs,
commonly referred to as {\em time-over-threshold}
and referred to as {\em pulse width} herein, 
was determined in the offline analysis and, being
a function of the pulse height, provided an estimate of
the charge deposited on each strip~\cite{hattori12}.
Measuring the pulse widths in this way allowed for a more
compact system 
and higher data rates than would be possible using conventional ADCs
for energy measurement.
However,
individual events had to be isolated from the data stream in the offline analysis.

\begin{figure}[ht]
\centering
\includegraphics[width=6.5 cm,clip]{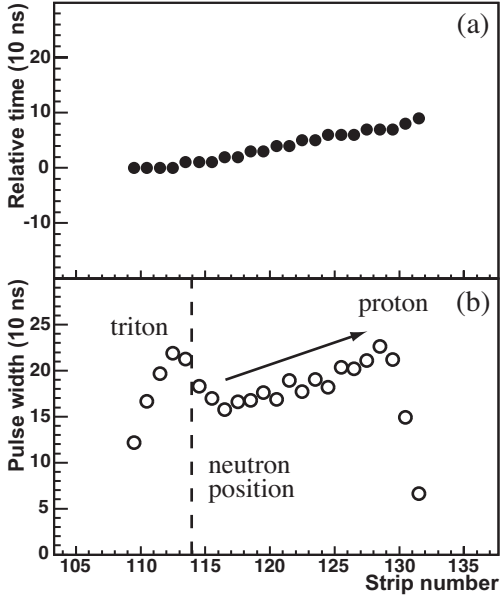}
\caption{\label{fig:pttrack} Proton-triton track measured with the
prototype neutron detector.
(a) Position (strip number) versus time and (b) pulse width for each hit strip
are shown.
The neutron interaction position (dashed line) and the slow rise of the
proton Bragg peak (arrow) are indicated in (b).
}
\end{figure}

\begin{figure}[ht]
\centering
\includegraphics[width=7.6 cm,clip]{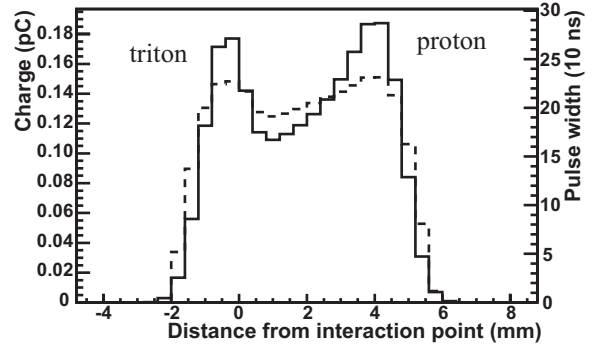}
\caption{\label{fig:ptsim} Simulation of pulse-width distributions.
The charge deposition (solid line) and pulse-width distribution 
(dashed line) are shown for a proton-triton track as determined by our 
GEANT4 simulation.
}
\end{figure}

Fig.~\ref{fig:pttrack} shows a neutron interaction event 
measured using our prototype system.
The position of the track, given by the relative time and strip number
of each hit, is
shown in Fig.~\ref{fig:pttrack}(a), with the associated pulse widths shown in 
Fig.~\ref{fig:pttrack}(b).
The observed pulse-width distribution clearly displays the Bragg peaks of 
the stopping proton and triton and shows good agreement with
the expected distribution, determined using our GEANT4 simulation and 
plotted in Fig.~\ref{fig:ptsim} (dashed histogram).
While tracking alone gives no indication of the neutron interaction point,
the proton and triton can be distinguished
from the shape of the pulse-width distribution.

\section{Event reconstruction}
\label{sec:ana}

The first task of the offline analysis was to cluster the raw hits
from the data stream produced by the FPGA encoder
into individual events.
This was accomplished in three steps:
1) match leading- and trailing-edge hit pairs and calculate pulse widths,
2) combine hits into anode and cathode clusters based 
on proximity in time and space, and
3) pair anode and cathode clusters based on proximity in time to make 
individual {\em events}.
Once the events were reconstructed, 
interesting quantities such as the track length
and pulse-width sum (for neutron identification) and
the shape of the pulse-width distribution (for proton-triton separation)
were evaluated as described below.
The offline data analysis was carried out using software based on
the ROOT object-oriented framework~\cite{root}.

\subsection{Track-length determination}
\label{sec:tl}

The 3-dimensional track length, $l$, was determined by combining the 2-dimensional
hit positions measured by the $\mu$PIC with the hit-time information as:

\begin{equation}
l = \sqrt{(\Delta x - x_0)^2 + (\Delta y - y_0)^2 + v_d^2 (\Delta t - t_0)^2},
\end{equation}
where $\Delta x$ and $\Delta y$ are 
the difference between the highest and lowest hit-strips for the anodes 
and cathodes, respectively, 
and $\Delta t$ is the difference in the time of the earliest and latest
hits which, when multiplied by the electron drift velocity, $v_d$,
gives the component of the track length perpendicular to the $\mu$PIC.
(The quantity $\Delta t$ was determined separately for anodes and cathodes 
and then averaged.)
The drift velocity was estimated using the MAGBOLTZ program~\cite{magboltz}
and had a value of $v_d \simeq 23$~$\mu$m/ns for the 2-atm Ar-C$_2$H$_6$-$^3$He
(63:7:30) gas mixture and 800-V/cm drift field used here.
The offsets $x_0$, $y_0$, and $t_0$ account for effects such as electron 
diffusion and decay time of the analog signals
and were taken from the minimum measured values of $\Delta x$, $\Delta y$, 
and $\Delta t$, respectively.

\begin{figure}[ht]
\centering
\includegraphics[width=7.5 cm,clip]{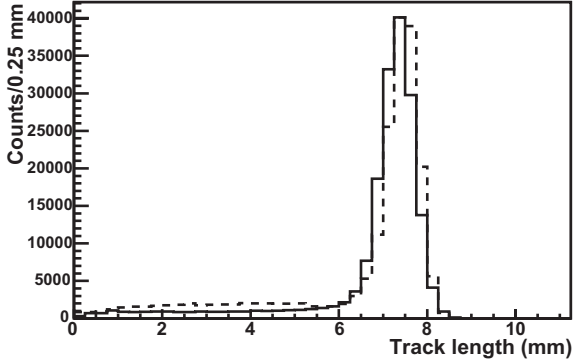}
\caption{\label{fig:trklen} Distribution of track lengths.
Track lengths for data taken with a moderated $^{252}$Cf source (solid)
are shown with the expected distribution for thermal neutrons given 
by GEANT4 (dashed).
The long tail is due to neutron events that were not fully contained in
the detector. }
\end{figure}

Fig.~\ref{fig:trklen} shows the resulting track-length distribution for 
neutrons from a $^{252}$Cf source  
moderated through loose-packed polyethylene pellets with an overall 
thickness of $\sim$10~cm.
The expected distribution 
for thermal neutrons determined from our GEANT4 simulation is overlaid as
the dashed histogram.
The observed spread in the track-lengths of the proton-triton events 
is the result of multiple scattering and electron
diffusion in the gas of the detector
coupled with the finite strip pitch and non-zero ASD thresholds.
The long tail extending to smaller track lengths 
is mainly due to neutron events that were not fully contained within 
the active volume of the detector.
The slight difference between the data and simulation is a result of
uncertainties in the gains of the $\mu$PIC and ASD amplifiers,
as well as the simplified model used to emulate the response of the ASDs
in our simulation.

A side effect of the streamed nature of the encoder data was that,
while the relative distance between hits within a track could be calculated
as described above,
the absolute position of the track above the surface of the $\mu$PIC 
could not be determined.
This was a direct result of the lack of a trigger reference time and 
lead to some uncertainty in the time for each event.
The minimum uncertainty was determined from the time required for an 
electron cloud to drift across the entire drift height and has a
value of about $\pm$0.6~$\mu$s
for the 2.5-cm drift cage and 
gas filling described in Sec.~\ref{sec:design}.
The event-time uncertainty then increases as the velocity of the 
incident neutron decreases
(to about 3~$\mu$s for thermal neutrons at 2200~m/s 
and up to 18~$\mu$s for 1-meV neutrons at 440~m/s,
as determined by simulation).
This effect dominates the timing resolution of our detector.

\subsection{Pulse-width sum}
\label{sec:pwpid}

The sum of pulse widths (or {\em total time-over-threshold})
was used as a rough estimate of total energy deposition.
This was possible since the pulse widths are
a function of the charge (and thus energy) deposited on each strip,
as shown in Ref.~\cite{hattori12}.
For the purposes of the present analysis,
it was not necessary to convert the widths to absolute charge or energy
(i.e., that the relationship exists was sufficient, so that the pulse widths could
take the place of energy deposition).
This pulse-width sum, along with the track length described above,
formed the basis for neutron identification.

\begin{figure}[ht]
\centering
\includegraphics[width=7.5 cm,clip]{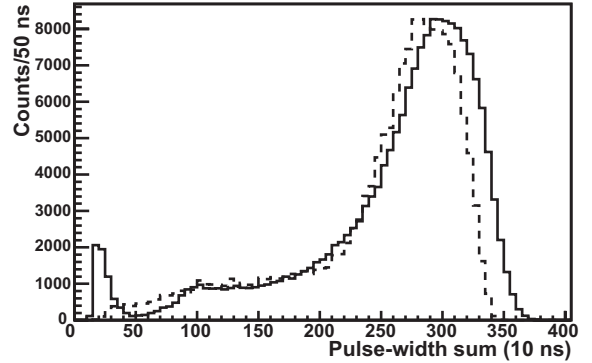}
\caption{\label{fig:pwsum} Pulse-width sum distribution.
The pulse-width sum distribution for neutrons from a $^{252}$Cf source (solid)
overlaid with the result of a GEANT4 simulation (dashed).
The additional small peak seen in the data is due to the gammas produced
along with the neutrons in $^{252}$Cf decay. }
\end{figure}

The distribution of the pulse-width sum for the data taken with
the moderated $^{252}$Cf source is shown in Fig.~\ref{fig:pwsum}.
The neutron events are clustered at higher values,
producing the large peak seen in the plot.
The small peak near the lower end is due to the gammas  
produced alongside the neutrons in the decay of $^{252}$Cf.
The large width of the neutron peak arises from a dependence of the 
measured pulse widths on the angle of the proton-triton track relative to
the $\mu$PIC,
and the long tail down to lower values is due to tracks that were not
fully contained within the active volume of the detector.
The data is in general agreement with the result of our GEANT4 simulation
for thermal neutrons (overlaid on Fig.~\ref{fig:pwsum} as the dashed histogram),
with slight differences due to the uncertainties in the gain and ASD
response as mentioned in the previous section.

\subsection{Proton-triton separation for position reconstruction}
\label{sec:ptsep}

Proton-triton separation, essential for the accurate reconstruction of
the neutron position, was carried out using the shape of the pulse-width
distribution.
As shown in Figs.~\ref{fig:pttrack}(b) and \ref{fig:ptsim},
the triton deposits most of its energy in a relatively short distance, 
while the proton produces a
gently sloping curve with the peak deposition near the stopping point
(giving the familiar Bragg curve).
The proton direction was then determined 
for the tracks which show this characteristic double-peaked
shape in the pulse-width distribution
by noting which end of the track
displays the relatively slow rise of the Bragg curve of the proton,
as indicated in Fig.~\ref{fig:pttrack}(b).

\begin{figure}[ht]
\centering
\includegraphics[width=8.5 cm,clip]{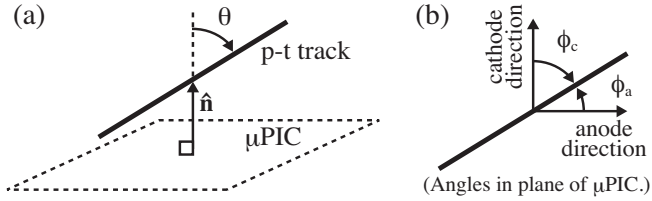}
\caption{\label{fig:dir} Definition of track orientation.
The track orientation for an anode or cathode cluster is given by two angles:
(a) the polar angle $\theta$ defined relative to the normal to the plane of
the $\mu$PIC, and 
(b) the azimuthal angle $\phi_a$ ($\phi_c$) defined relative 
to the anode (cathode) strip direction.
All angles are restricted to the range 0 to 90 degrees.
}
\end{figure}

Using our GEANT4 simulation, the efficiency of the proton-triton
separation and determination of the proton direction,
referred to as the PTS (proton-triton separation) efficiency, 
was estimated at $\sim$95\%
for tracks with angles $\theta > 60^\circ$ and $\phi_{a,c} > 30^\circ$
(for $\theta$ and $\phi_{a,c}$ as defined in Fig.~\ref{fig:dir}).
As the PTS procedure was performed for both the anode and cathode clusters,
the PTS efficiency {\em per neutron event} becomes $\sim$90\%.
While the efficiency was roughly constant over the above range of angles,
it dropped to zero for track angles $\theta < 20^\circ$ 
or $\phi_{a,c} < 10^\circ$.
This was due to the fact that, as one goes to smaller track angles, 
the electron cloud becomes concentrated over fewer and fewer strips,
making it increasingly difficult to observe the double-peak structure.
This was simply a limitation of the 2-dimensional strip readout and
time-over-threshold methods used here.

After applying the PTS procedure,
the corrected neutron interaction position, $(X',Y')$,
was calculated as:

\begin{align}
\label{eqn:pcor}
X' & = X - \hat{x} \left(f_p - \tfrac{1}{2}\right) (\Delta x - x_0), \nonumber \\
Y' & = Y - \hat{y} \left(f_p - \tfrac{1}{2}\right) (\Delta y - y_0),
\end{align}
where $X$ ($Y$) is the mid-point of the proton-triton track
for the anodes (cathodes) and
$\hat{x}$ and $\hat{y}$ indicate the proton direction and take a value of $+1$
($-1$) for a proton moving toward higher (lower) strip numbers
or $0$ if the proton and triton were not visible. 
The correction factor $f_p = 0.785$ is the ratio of the proton track length to
the total proton-triton track length 
and was determined using our GEANT4 simulation.
The remaining terms give the $x$ and $y$ component of the total track length 
as described in Sec.~\ref{sec:tl}.
Neutron events for which the proton and triton could not be seen in either
cluster ($\sim$15\% of all events) should give the worst spatial resolution,
while events in which they were visible in both clusters 
($\sim$35\%) should give the best.
The improvement in the spatial resolution achieved by this method 
will be discussed in Sec.~\ref{sec:pres}.

\begin{figure}[ht]
\centering
\includegraphics[width=6.56 cm,clip]{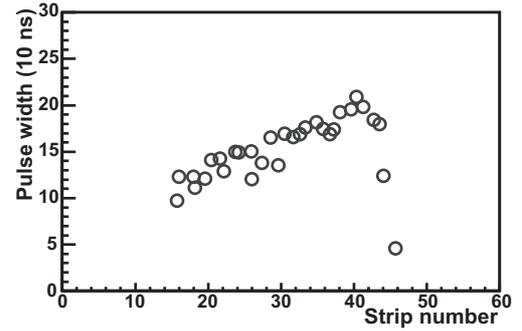}
\caption{\label{fig:fastn} Scattered proton track measured with
the prototype detector.
Pulse width distribution for a proton scattered from the quencher gas 
by a fast neutron.
}
\end{figure}

Another benefit of measuring the pulse-width distributions in this way
relates to the background caused by fast neutrons (i.e., 
neutrons with energy on the order of MeV).
These fast neutrons sometimes scatter with protons present in the
materials of the detector, with
a small fraction
producing tracks with similar length and total energy deposit as the
proton-triton tracks.
The pulse-width distribution shown in Fig.~\ref{fig:fastn} is
one such scattered proton track, as measured with our prototype detector.
Possessing no triton Bragg peak,
the shape of the distribution is quite distinct from that of the proton-triton
track of Fig.~\ref{fig:pttrack}(b), 
providing the only means to reject such events.

\section{Results and discussion}

\subsection{Neutron-gamma separation}
\label{sec:ngsep}

\begin{figure}[ht]
\centering
\includegraphics[width=7. cm,clip]{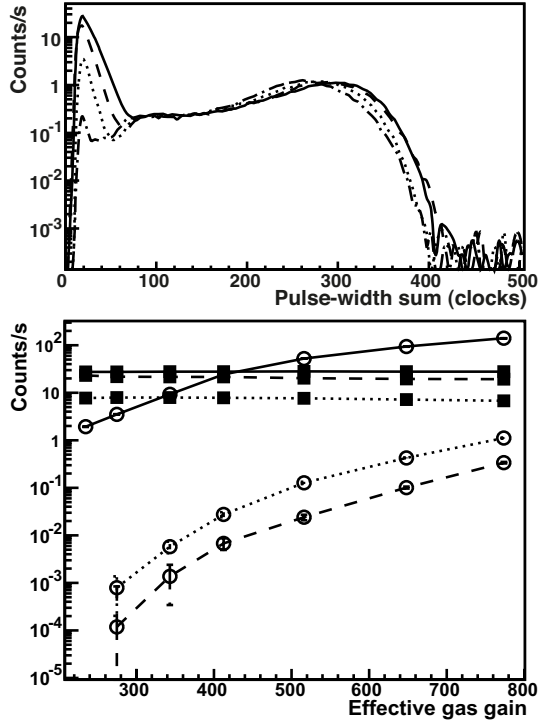}
\caption{\label{fig:cpsvg} Neutron and gamma event rates as a function
of gas gain. 
The upper plot shows the pulse-width sum 
at gain settings of 650 (solid line), 510 (dashed line), 345 (dotted line),
and 230 (dash-dotted line).
In the lower plot,
the numbers of neutrons (squares) and gammas (open circles) detected 
per second are shown for: 1) no tracking
cuts (solid lines), 2) with a track length cut (dashed lines), and 3)
with the PTS cut (dotted lines) as described in the text.
}
\end{figure}

Since the energy deposition for gamma events was near our detection threshold,
the gamma detection efficiency, or the probability a gamma will interact
in the detector and deposit enough energy in the active volume to produce data, 
was sensitive to the gain setting of the $\mu$PIC.
This was in contrast to neutron events, which, being well above threshold,
had a detection efficiency with virtually no gain dependence.
To demonstrate this,
the detector was irradiated simultaneously with a 1-MBq $^{137}$Cs gamma source
placed in the center of the entrance window
(giving an effective dose-rate of about 16~$\mu$Gy/h over the active volume)
and a $^{252}$Cf source, emitting $\sim$$2 \times 10^4$~neutrons/s, 
located about 14~cm directly above the center of the entrance window, 
and moderated through $\sim$10~cm of polyethylene 
(consisting of pellets of size $\sim$4~mm).
The detector was outfitted with a 2.5-cm drift cage and
filled with Ar-C$_2$H$_6$-$^3$He (63:7:30) at 2~atm.
The gas gain was controlled by altering the voltage applied to the anodes of
the $\mu$PIC, and data was taken for 45 minutes at each gain setting.
The results, normalized to counts per second, 
are shown in Fig.~\ref{fig:cpsvg}.

The upper plot of Fig.~\ref{fig:cpsvg} shows the pulse-width sum distributions
for several different gain settings, and the lower plot gives
the corresponding neutron (squares) and gamma (open circles) rates.
For the lower plot of Fig.~\ref{fig:cpsvg},
the neutrons and gammas were separated by an `energy' cut in the pulse-width sum,
with events above the cut considered neutrons and those below considered gammas.
The total gamma rate 
decreased by nearly two orders of magnitude between the highest and lowest
gain settings, while the total neutron rate remained virtually constant
(total rates indicated by solid lines).
Also shown are the effective neutron and gamma event rates
after applying a track-length cut of $-$2$\sigma$ to $+$3$\sigma$ (dashed lines),
where $\sigma$ is the Gaussian width of the track-length
distribution of Fig.~\ref{fig:trklen},
or after applying a PTS cut
requiring the separation of the proton and triton in both 
the anode and cathode clusters of each event (dotted lines).
These tracking cuts resulted in even larger reductions in the gamma event rate.

\begin{figure}[ht]
\centering
\includegraphics[width=8.5 cm,clip]{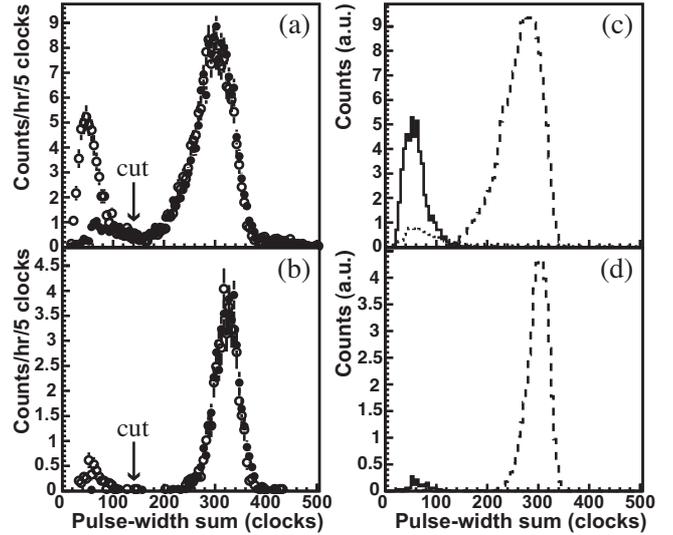}
\caption{\label{fig:ngsep} Neutron-gamma separation.
Pulse-width sums for events from a $^{137}$Cs
source (open circles) and with no source (closed circles) 
are shown for (a) track-length cut only and (b) track-length and PTS cuts.
Plots (c) and (d) show simulated data, including 662-keV gammas (solid line), 
neutrons (dashed line), and cosmic-ray muons (dotted line),
for the same cut conditions as (a) and (b), respectively.
The `gamma cut' at 140 clocks is also indicated. 
}
\end{figure}

To further quantify the effectiveness of gamma rejection via track
properties, more data was taken with the $^{137}$Cs source at a gain
setting of $\sim$600.
(This gain setting provided a high enough count rate such that the
measurement could be completed in a reasonable time and 
is similar to the gain setting used for the measurement of Sec.~\ref{sec:pres}.)
Using the same detector setup and gas filling as above,
data was collected for 24 and 48 hours with and without the source,
respectively.
The resulting pulse-width sum distributions, normalized to counts per hour, 
are shown in Fig.~\ref{fig:ngsep}(a)
after the $-$2/$+$3$\sigma$ track-length cut and Fig.~\ref{fig:ngsep}(b)
for both the track-length and PTS cuts.
The distributions from $^{137}$Cs (open circles) are overlaid with the
no-source distributions (closed circles).
The large peak is due to thermal neutrons produced by cosmic rays, while
the gammas from the $^{137}$Cs source are 
clearly visible as the smaller peak near the lower part of the distributions.
The plots in Figs.~\ref{fig:ngsep}(c) and (d) show the results from our 
GEANT4 simulation for neutron, gamma, and cosmic-ray muon events for the same
cuts as Figs.~\ref{fig:ngsep}(a) and (b), respectively.
The simulated data reproduces the features seen in the real data quite well.

The gamma contamination for a given set of tracking cuts was found 
by integrating the pulse-width sum distributions above the `energy' cutoff
indicated in Fig.~\ref{fig:ngsep}
and subtracting the number of background counts observed for the 
same conditions. 
The resulting negative numbers, as shown in Table~\ref{tbl:gcont},
are due to variations in the background cosmic-ray rate, 
limited statistics due to the overall low background rate,
and slight variations in the $\mu$PIC gain over time.
In light of this, the final results are given as upper limits 
determined in a way that accounts for the 
physical constraint that the rates cannot be less than zero~\cite{fc98}.
After normalizing to the total number of gammas detected per hour, 
gamma contamination fractions
of $< 9.3 \times 10^{-6}$ and $< 3.6 \times 10^{-6}$ (95\% confidence level) were 
found for the data shown in Figs.~\ref{fig:ngsep}(a) and \ref{fig:ngsep}(b),
respectively.
Taking into account the gamma detection efficiency at the above gain setting
($\sim$$6 \times 10^{-4}$),
a gamma sensitivity on the order of 10$^{-9}$ or less was determined
for our prototype detector when operated at a gain of $\sim$600.
Furthermore, from
the results of Fig.~\ref{fig:cpsvg} for the gamma rates after track
length and PTS cuts, we estimated that 
a decrease in the gas gain to around 275 would reduce the gamma
sensitivity by another three orders of magnitude, 
resulting in an effective gamma sensitivity on the order of $10^{-12}$ or less
with no change in neutron efficiency.

\begin{table*}[ht]
\caption{\label{tbl:gcont} Gamma rejection with tracking cuts
for irradiation with a 1-MBq $^{137}$Cs source.
Data are given in counts/hour.
The contamination fraction is the ratio of the contamination rate 
for a given set of cuts to the contamination rate with no cuts.
Upper limits (given in parentheses) are at the 95\% confidence level.}
\begin{center}
\tabcolsep 5.8pt
\small
\begin{tabular}{lrrrr}
\hline\hline
Cuts & \multicolumn{1}{c}{$^{137}$Cs} & \multicolumn{1}{c}{No source} 
 & \multicolumn{1}{c}{Contamination} 
 & \multicolumn{1}{c}{Contamination fraction} \\
\hline
None & $351360 \pm 120$ & $3700.4 \pm 8.8$ & $347660 \pm 120$ & $1$ \\ 
Track length & $185.0 \pm 2.7$ & $161.1 \pm 1.8$ & $23.9 \pm 3.3$ 
 & $6.87 \times 10^{-5}$ \\
Energy & $284.2 \pm 3.4$ & $290.1 \pm 2.5$ & $-5.8 \pm 4.2$ 
 ($<3.65$) & ($<1.0 \times 10^{-5}$) \\ 
Track length $\wedge$ Energy & $145.6 \pm 2.4$ & $148.7 \pm 1.8$ & $-3.2 \pm 3.0$ 
 ($<3.22$) & ($<9.3 \times 10^{-6}$) \\ 
Track length $\wedge$ Energy $\wedge$ PTS & $43.4 \pm 1.3$ & $46.1 \pm 1.0$ 
 & $-2.7 \pm 1.7$ ($<1.24$) & ($<3.6 \times 10^{-6}$) \\ 
\hline\hline
\end{tabular}
\end{center}
\end{table*}

For comparison, we consider $^3$He tubes, which, 
while not suitable for high-resolution imaging,
provide the best neutron-gamma separation among commonly used
detectors.
Owing to the detailed tracking capabilities of our detector,
the effective gamma sensitivity found above is on par or 
better than that of $^3$He tubes (see for instance, Ref.~\cite{kouzes11}),
which separate neutrons and gammas via pulse-height only.
(Due to differing experimental conditions and detector geometries, 
it is difficult to make a direct comparison.)
Furthermore,
the neutron identification efficiency for a $^3$He tube
can have a strong dependence on the operating voltage,
which must be optimized to obtain acceptable
gamma rejection at a given background rate
(see Ref.~\cite{beddingfield00} for typical $^3$He-tube plateau curves).
This is in contrast to the present results of Fig.~\ref{fig:cpsvg} (lower plot),
which show excellent stability for neutron identification versus gas gain.

\subsection{Spatial resolution}
\label{sec:pres}

The spatial resolution of the prototype detector 
was studied using a cadmium test chart.
The data were taken in November 2009 at the Japan Spallation 
Neutron Source (JSNS) located at the Japan Proton Accelerator
Research Facility (J-PARC) in Tokai, Japan.
The measurements were carried out at NOBORU (NeutrOn Beamline for 
Observation and Research Use)~\cite{noboru09}, 
with a beam power at the time of the experiment of $\sim$100~kW.
The prototype detector was filled with the same gas mixture as in the
previous section and operated at a gas gain around 600,
and a 5-cm drift cage was installed above the $\mu$PIC.
The detector was placed in the beam at a distance of 14.5~m from the
moderator, and the cadmium test chart, of size $50 \times 50$~mm$^2$, 
was attached directly to the entrance window.

\begin{figure}[ht]
\centering
\includegraphics[width=6. cm,clip]{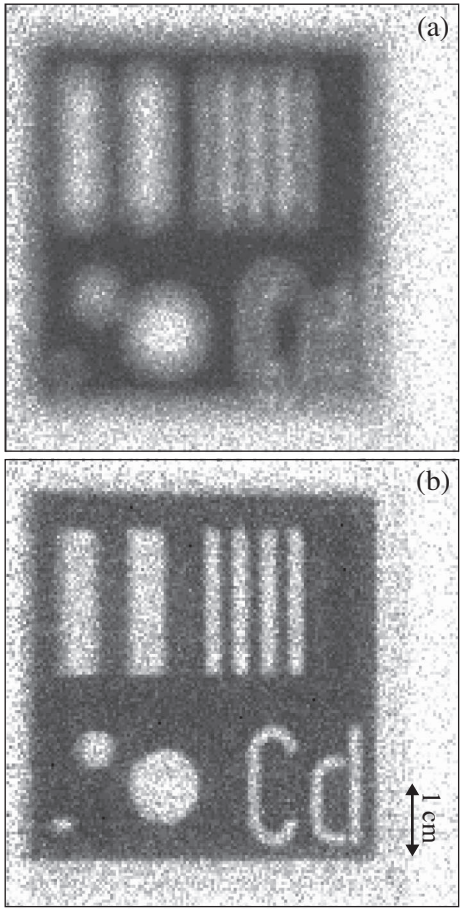}
\caption{\label{fig:cd} 2-dimensional images of a cadmium test chart.
(a) Image with neutron positions determined as the mid-point of the 
proton-triton tracks.
(b) Image with corrected neutron positions after proton-triton separation.
Lighter areas indicate higher neutron transmission.
}
\end{figure}

The resulting images, shown in Fig.~\ref{fig:cd}, were taken with an exposure time
of 9.2 minutes at a neutron rate of $\sim$42~kcps.
The numbers of events remaining after each tracking cut are listed in 
Table~\ref{tbl:cddata}.
The TOF cut (equivalent to a neutron energy cut of $E_n < 0.12$~eV)
selected neutrons with energy well below the cadmium cut-off, and
the looser energy and track length cuts, as compared to
the previous section, were possible due to the minimal gamma contamination
in the selected TOF region.
The final images were normalized to the beam profile measured with no
sample present (at similar statistics).

\begin{table}[ht]
\caption{\label{tbl:cddata} Data for image of Cd test chart.
The number of beam pulses and total number of events included in the
present analysis are given.
Also shown are the numbers of events remaining after each
subsequent tracking cut.
}
\begin{center}
\begin{tabular}{lrrc}
\hline\hline
 & Number & Fraction & Note \\
\hline
Beam pulses & 13871 & $-$ & \\
Events & $2.35 \times 10^7$ & $-$ & \\
${\rm TOF} > 3$~ms & $3.38 \times 10^6$ & 1 & \\
${\rm Energy} > 50$~clocks & $3.35 \times 10^6$ & 0.99 & \\
Track length ($\pm 3\sigma$) & $2.85 \times 10^6$ & 0.84 & Fig.~\ref{fig:cd}(a) \\
PTS cut & $9.90 \times 10^5$ & 0.29 & Fig.~\ref{fig:cd}(b) \\
\hline\hline
\end{tabular}
\end{center}
\end{table}

For the image in Fig.~\ref{fig:cd}(a), the pulse-width information was disregarded,
and the neutron position was taken as the mid-point of the proton-triton
track ($X$, $Y$ in Eq.~\ref{eqn:pcor}).
The resolution was estimated by fitting the projection of a straight edge
(sometimes referred to as a {\em knife-edge target})
with the function: 

\begin{equation}
\label{eqn:fit}
N(x) = \frac{A}{2} \left[ 1 - {\rm erf}\left(\frac{x - \mu}{\sqrt{2} \sigma}\right)\right] + C,
\end{equation}
derived assuming a Gaussian resolution,
where $A$ is a normalization factor, erf$()$ is the usual Gaussian error function,
$\mu$ is the mean position of the edge,
$\sigma$ is the corresponding Gaussian width,
and $C$ is a constant.
Using the right-most vertical edge of the test chart, the resolution was 
found as $\sigma = 990 \pm 180$~$\mu$m. 
This is in good agreement with a previous test experiment carried out
by our group in September 2008
at the NOP beamline of the JRR-3 reactor in Tokai, Japan, 
using a $\mu$PIC-based X-ray
imaging detector modified to detect neutrons (by adding a small amount 
of $^3$He to the usual gas mixture)~\cite{hattori_thesis}.
This previous experiment,
which used an older version of the FPGA firmware
without the time-over-threshold measurement capability, 
found a spatial resolution of about 956~$\mu$m.
(The $\mu$PIC and DAQ hardware were essentially identical to that used 
for the current prototype.)

Using the pulse-width information available 
in the current data,
the image quality was greatly improved (Fig.~\ref{fig:cd}(b))
by separating the proton and triton and calculating the
neutron interaction points via Eq.~\ref{eqn:pcor}. 
The spatial resolution was again determined via Eq.~\ref{eqn:fit}
and was found to be $\sigma = 349 \pm 36$~$\mu$m, 
representing an improvement by a factor of 2.8.
To achieve this spatial resolution, the events were
limited to those in which the proton and triton were visible in both the
anode and cathode clusters (PTS cut).
As discussed in Sec.~\ref{sec:ptsep} (and confirmed in Table~\ref{tbl:cddata}),
such events account for only about 35\% of all neutron events,
resulting in a combined neutron efficiency for the image of $\sim$10\% 
for thermal neutrons. 
When all events passing the energy and track-length cuts 
are included in the image, the spatial resolution becomes $523 \pm 39$~$\mu$m, 
which, although still an improvement over the case with no position correction,
is about 50\% worse than the best resolution.
Thus, a trade-off must be made between resolution and efficiency.

The distortion apparent on the left side of Fig.~\ref{fig:cd}(b) is a result of
beam divergence, which produces a smearing of the image that
increases with the distance from the beam center. 
As with the time resolution discussed in Sec.~\ref{sec:tl},
this was a result of our inability to measure the absolute position 
of an event above the surface of the $\mu$PIC.
While the effect is clearly visible on the left of the image, it
was estimated via simulation 
to increase the observed spatial resolution by only about 1\%
at the edge used in the above determination 
(well within errors).
This smearing can be reduced by optimizing the beam geometry to 
minimize divergence and by decreasing the drift height to
reduce the uncertainty in the interaction position.
The latter point led us to change from the 5-cm drift cage
to the 2.5-cm cage used in previous sections.

Improvement of the spatial resolution can be attained with more sophisticated
position reconstruction algorithms incorporating the shape of the
pulse-width distribution
(as in Ref.~\cite{hattori12} for X-ray imaging with the $\mu$PIC).
Preliminary results for our neutron detector suggest nearly a factor of 3 
improvement in the resolution is possible~\cite{parker11}.
Further improvement is expected through optimization of the gas mixture
and pressure,
although, according to our GEANT4 simulation, 
the spatial resolution is ultimately limited to between 90 and 100~$\mu$m by the 
underlying pixel-pitch.
Our simulation showed that this maximal value
is achieved for a gas mixture with a diffusion constant about half 
the current mixture or one with a track length around 5~mm.
For the latter,
shorter track lengths (down to a minimum of $\sim$3~mm for which the
proton and triton can be effectively separated) would reduce the amount of
data per event but would also yield a smaller
range of usable track angles and a reduced reconstruction efficiency.
A trade-off is thus affected
between spatial resolution, efficiency, and maximum neutron rate.
These ongoing studies will be explored in a subsequent paper.

\subsection{Long-term operability and $^3$He usage}

At a time when the demand for $^3$He is already outstripping
the world's supply~\cite{he3supply},
it is important to consider the $^3$He usage of the detector.
Over time, various processes, 
such as dissociation of the quencher gas 
and outgassing from the materials of the detector,
lead to a buildup of impurities and a degradation in the 
properties of the gas.
To maintain the desired gas properties, the detector must be periodically
evacuated and refilled.
To date,
the prototype detector has been filled two times with the gas mixture
described in this paper.
The first filling was used for a period of 13 months, 
and the second has been used up to the present ($\sim$14 months).
The gas gain slowly diminished over each fill period, 
decreasing by 68\% during the first and by 30\% during the second.
Also, the initial value of the second filling was 56\% that of the first, 
perhaps indicating some additional aging process of the $\mu$PIC.
In each case, however,
sufficient gain for neutron detection could still be attained by increasing
the anode voltage setting,
with no significant change observed in detector performance over time. 

A one-year fill cycle is already long when compared to other
$\mu$PIC applications, which, 
due to the need to operate at much higher gains,
must be refilled every few months.
By treating the vessel and $\mu$PIC against outgassing through annealing
(not done for our prototype)
and employing a gas filtration system, such as one now under development in our 
group at Kyoto University~\cite{gasfilter},
it should be possible to stabilize the gas gains over time
and extend the operation of the detector significantly.
If these improvements can be realized,
the yearly costs for detector maintenance would be greatly reduced.

\section{Conclusion}

We have developed a time-resolved neutron imaging detector based on 
the $\mu$PIC micro-pixel chamber.
By measuring the energy deposition (via time-over-threshold) 
and track positions,
we were able to demonstrate
a small effective gamma sensitivity of ${\scriptstyle \lesssim}$10$^{-12}$ and
neutron identification that was stable and robust.
A spatial resolution of $\sim$350~$\mu$m was achieved, with significant 
improvements expected through the use of more sophisticated reconstruction 
algorithms and optimization of the gas mixture.
A time resolution as low as 0.6~$\mu$s allows for clean separation of
neutrons by time-of-flight at pulsed neutron sources.
The detector, operated at a moderate pressure of a few atmospheres,
was able to maintain its good operating characteristics for more than
one year on a single gas filling.
The compact FPGA-based data acquisition system 
is simple to setup and operate and,
although limited to a maximum neutron rate of $\sim$150~kcps,
could potentially support a rate 10 times that
by upgrading the encoder-to-PC data transfer method.
Furthermore,
the printed-circuit board manufacturing process is inexpensive and
allows some freedom to
choose the size and shape of the $\mu$PIC to suit a particular experiment
(up to a maximum size of $30 \times 30$~cm$^2$), and
larger areas can be covered by tiling detectors with no degradation
in the operating characteristics with increased area.
These properties make the $\mu$PIC an attractive solution for neutron
imaging applications requiring very low background, time-resolved
measurements, 
and the ability to cover large areas at moderate resolution.

\section*{Acknowledgements}

This work was supported by the Quantum Beam Technology Program of the Japan 
Ministry of Education, Culture, Sports, Science and Technology (MEXT).
The authors would like to thank the staff at J-PARC and the 
Materials and Life Science Facility for
providing a stable neutron beam during our test experiment in November 2009.

%%
%%%%%%%%%%% end of main text %%%%%%%%%%%%%%%%%%



%% References


\bibliographystyle{elsarticle/model1-num-names}
%\bibliography{references.bib}
%\end{document}



\end{document}